\documentclass[12pt,preprint]{aastex}
\usepackage{natbib,epsfig}
\tighten

\newlength{\tskip}
\newlength{\colwidth}

\newcommand{\beq}{\begin{equation}}
\newcommand{\eeq}{\end{equation}}
\newcommand{\beqa}{\begin{eqnarray}}
\newcommand{\eeqa}{\end{eqnarray}}

\slugcomment{{\it Astrophysical Journal} in press}

\begin{document}
\title{Contribution to Unresolved Infrared Fluctuations from Dwarf Galaxies at Redshifts of $2-3$}
\author{Ranga-Ram Chary\altaffilmark{1},
Asantha Cooray\altaffilmark{2}, 
Ian Sullivan\altaffilmark{1}}
\altaffiltext{1}{Division of Physics, Mathematics, and Astronomy,  
California Institute of Technology, Pasadena, CA 91125. Email: rchary@caltech.edu}
\altaffiltext{2}{Center for Cosmology, Department of Physics and  
Astronomy, University of California, Irvine, CA 92697. E-mail:  
acooray@uci.edu}
\begin{abstract}
In order to understand the origin of clustered anisotropies detected in {\it  
Spitzer} images between 3.6 and 8\,$\mu$m, we stack the {\it Spitzer} IRAC/Great Observatories
Origins Deep Survey (GOODS) images at pixel locations corresponding to
faint, $z_{AB}\sim27$ mag, optical sources with no obvious IR counterparts. We obtain a
strong detection of the sources
with a stacked median flux at 3.6$\mu$m of 130$\pm$5 nJy above the background. 
The wealth of multi-wavelength imaging data in GOODS enables a similar stacking analysis to be undertaken
at various wavelengths between the ultraviolet and near-infrared bands. We obtain strong stacked detections of these
optically faint sources over the entire wavelength range 
which places constraints on the average properties of these sources.  We find that
the flux spectrum of the median, stacked source
is consistent with a $L\lesssim0.03~L_{*,{\rm UV}}$ galaxy with a 90\% confidence interval for the redshift of $1.9-2.7$.
These sources produce a 3.6 $\mu$m absolute background intensity between 0.1 and 0.35 nW m$^{-2}$ sr$^{-1}$
and the clustered IR light could account for $\sim30-50$\% of fluctuation power in the IR background
at 4 arcminute angular scales. Although the exact redshift distribution of these sources is unknown, these galaxies
appear to contain $5-20$\% of the co-moving stellar mass density at $z\sim2.5$. 
\end{abstract}


\section{Introduction}
The intensity of the cosmic near-infrared background (IRB) is a  
measure of the total light emitted by stars and galaxies in the universe which is not  
thermally reprocessed by dust. 
The absolute background has been estimated with the
Diffuse Infrared Background Experiment (DIRBE; Hauser  
\& Dwek 2001) and the Infra-Red Telescope in Space (IRTS; Matsumoto et al. 2005)
which results in values that are about a factor of two higher than the
intensity obtained by integrating the light from individually detected galaxies.
This is most likely attributable to the large uncertainties associated with the removal of
foreground zodiacal light from the DIRBE observations (Dwek et al. 2005). 
This hypothesis has been strengthened by the TeV spectrum of gamma-ray blazars
which indicate that the total IRB intensity is smaller than the DIRBE/IRTS estimates
(Aharonian et al. 2005) and that $\sim$ 90\% of the IRB light probably arises in
known galaxy populations.
Yet, attempts have been made to explain
the difference between the measured and resolved IRB intensity 
with sources at the epoch of reionization (e.g., Santos, Bromm \& Kamionkowski 2002; Salvaterra \& Ferrara  
2003; Cooray \& Yoshida 2004; Fernandez \& Komatsu 2006). If first-light galaxies are to explain   the missing
intensity completely, then an extreme scenario is needed with the conversion of at least 5\% of  
all baryons to stars (Madau \& Silk 2004). 

Instead of the absolute intensity, recent works 
have concentrated on spatial   fluctuations of the IRB 
(Cooray et al. 2004; Kashlinsky et al. 2004).
While an interpretation of unresolved fluctuations is subject to
extremely uncertain astrophysical modeling of underlying faint populations,
fluctuations in deep {\it Spitzer}  images have  been fully attributed to first-light galaxies containing 
Pop~III stars during reionization (Kashlinsky et al. 2005). An alternative study
shows that a reasonable fraction ($>50$\%) of unresolved fluctuations is arising from
faint, unresolved sources at lower redshifts (Cooray et al. 2007). Thus, while 
two independent studies find a  similar clustering amplitude for
IRB fluctuations (Kashlinsky et al. 2004; Cooray et al. 2007), they  
differ in the interpretation related to the contribution from Pop~III stars due to differences in the background 
light ascribed to faint foreground galaxies unresolved by {\it Spitzer}.

The suggestion for a low redshift origin in Cooray et al. (2007)  comes from predictions using a halo model (Cooray \& Sheth 2002) 
for the 3.6 $\mu$m population matched to 
their luminosity functions  (Babbedge et al. 2006) and clustering power spectra (Sullivan et  al. 2007).
In Cooray et al. (2007) this low-redshift population was identified
as the faint optical galaxies that are resolved in the {\it Hubble} Advanced Camera for Surveys (ACS; Giavalisco et al.  
2004) images of the Great Observatories Origins Deep Survey (GOODS; Dickinson et al. 2003) fields, 
but not detected in  deep {\it Spitzer} Infrared Array Camera (IRAC)   images  of the same
fields. With models, it was estimated  that this faint galaxy population  has a 3.6 $\mu$m  intensity between  
0.1 and 0.8 nW m$^{-2}$ sr$^{-1}$. The analysis in Cooray et al. (2007) did not exclude a high redshift
contribution to the IRB, with an  upper  limit on the 3.6 $\mu$m absolute background intensity from
  $z>8$ sources of  0.6   nW m$^{-2}$ sr$^{-1}$. 
In comparison, if all unresolved IR fluctuations are generated by faint $z>6.5$ sources which are below the detection
threshold of {\it Spitzer}, then the 
IRB intensity for such sources is $\gtrsim$1 nW m$^{-2}$ sr$^{-1}$ at 3.6 $\mu$m (Kashlinsky et al. 2007c).

In this {\it Letter}, we further study the physical origin of fluctuations  
in the IR background. Since faint ACS optical sources are undetected
in deep {\it Spitzer} IRAC images, we stack their pixel locations in IRAC to establish
the average IR flux. We also utilize the wealth of multiwavelength imaging 
data in the GOODS fields to obtain a stacked broadband
spectral energy distribution of these faint optical galaxies between UV and near-IR bands.
Leaving the redshift as a free parameter, we fit the average optical to IR flux spectrum 
with galaxy population synthesis models and find that the stacked flux spectrum is best fit
by a $\sim0.03$~L$_{*, {\rm UV}}$ galaxy at $1.9<z<2.7$,
similar to a scaled-down population of Lyman-break galaxies.
We also measure the expected IRB fluctuation clustering spectrum  
from these sources. 

We  
summarize the stacking analysis in the next section
and discuss results on the average optical to IR flux spectrum
in \S~3. We discuss clustering of these faint optical  
sources in \S~4. In \S~5 we discuss the implications of our measurements.
The magnitudes quoted throughout this paper are
AB magnitudes.

\begin{figure}[!t]
\centerline{\psfig{file=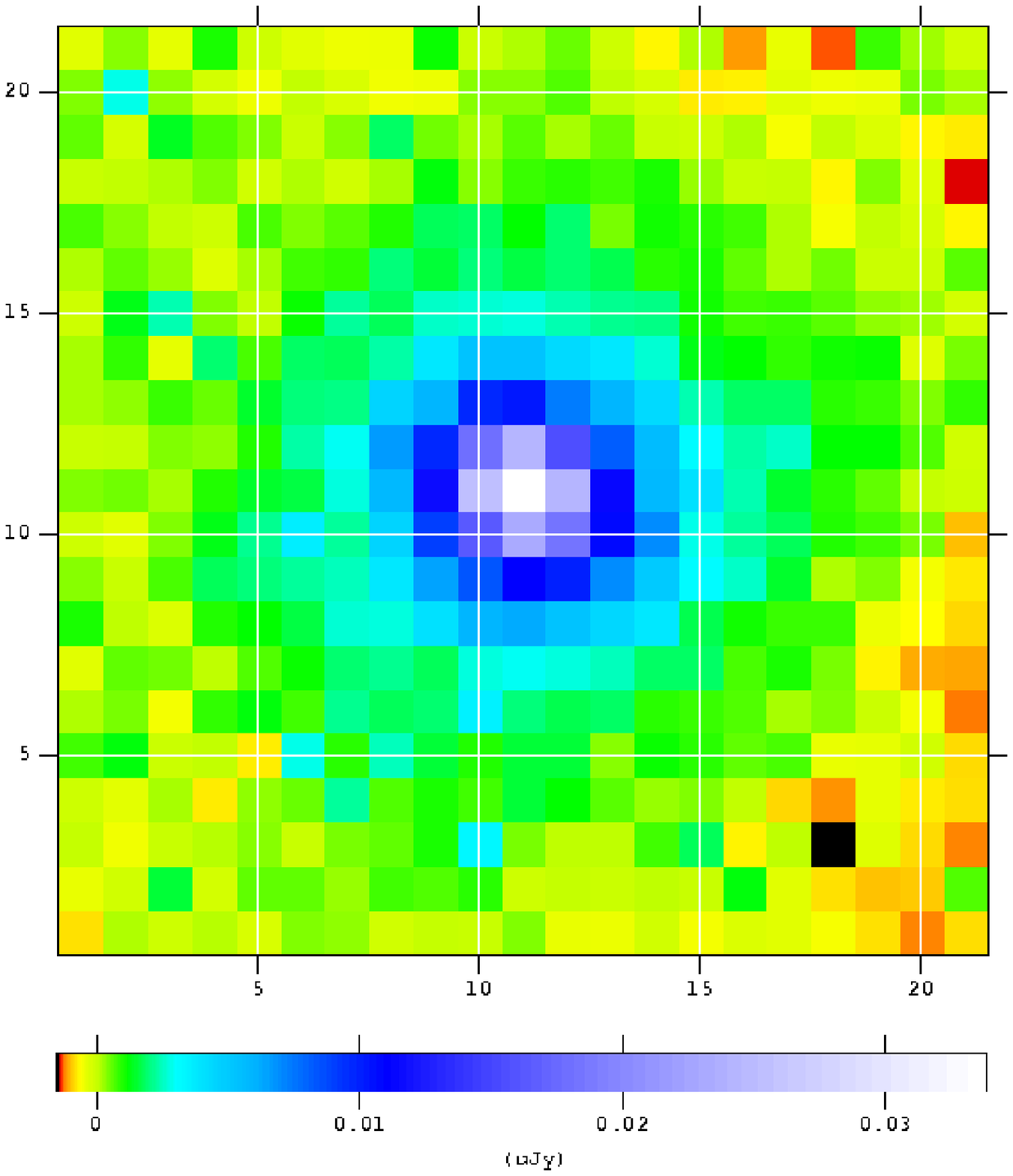,width=3.3in,angle=0}    }
\caption{
The average-stacked 3.6 $\mu$m image from GOODS-N with
pixel locations of all 6160 ACS sources unmatched and unmasked in IRAC.
The median stacked flux is  $(130 \pm 5)$ nJy (or $=26.1 \pm 0.1$ magnitude) 
and is detected with a  
statistical signal-to-noise ratio higher than 25. 
}
\label{fig:stack}
\end{figure}

\begin{figure}[!t]
\centerline{\psfig{file=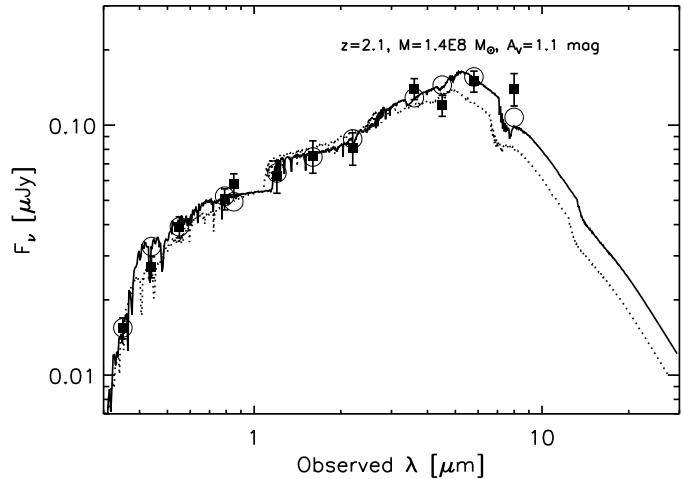,width=3.9in,angle=0}    }
\caption{
The stacked ultraviolet to IR flux spectral energy distribution
of faint galaxies detected by ACS but not detected individually by 
{\it Spitzer}/IRAC in GOODS-S (black line and data points). The SED is consistent 
with the sub-L$_{*,{\rm UV}}$ 
galaxy population at $z\sim2.5$. The parameters of the stacked galaxy
for GOODS-S are as shown in the legend. 
The convolution of the fit SED with the different passbands is shown with empty circles.
The stacking in GOODS-N results in a best
fit redshift of $z\sim1.9$ and the SED shown as the dotted line (See text for details). 
}
\label{fig:spec}
\end{figure}

\begin{deluxetable}{ccc}
\tablecaption{Results from Stacking Analysis\label{table:data}}
\tablehead{
\colhead{} &
\colhead{GOODS-N} & 
\colhead{GOODS-S} \\
}
\startdata
$\Delta \Omega$/arcmin$^2$ &   169  & 171 \\
N$_{\it ACS}$ &  6160 & 5441 \\
U & 28.3 $\pm$ 0.1 & 28.4 $\pm$ 0.1 \\
B & 28.0 $\pm$ 0.1 & 27.8 $\pm$ 0.1 \\
V & 27.5 $\pm$ 0.1 & 27.4 $\pm$ 0.1 \\
i & 27.2 $\pm$ 0.1 & 27.1 $\pm$ 0.1 \\
z & 27.1 $\pm$ 0.1 & 27.0 $\pm$ 0.1 \\
HK$\arcmin$ & 27.3 $\pm$ 0.3 & ... \\
J & ... & 26.9 $\pm$ 0.15 \\
H & ...	& 26.7 $\pm$ 0.15 \\
Ks & ... & 26.6 $\pm$ 0.15 \\
3.6 $\mu$m  & 26.1 $\pm$ 0.1 &  26.0 $\pm$ 0.1 \\
4.5 $\mu$m & 26.2 $\pm$ 0.1 & 26.2 $\pm$ 0.1 \\
5.8 $\mu$m & 26.2 $\pm$ 0.1 & 26.0 $\pm$ 0.1 \\
8.0 $\mu$m & 26.3 $\pm$ 0.15 & 26.0 $\pm$ 0.15 \\
\enddata
\tablecomments{N$_{\it ACS}$ is the total number of faint ACS sources over an  
effective area
$\Delta \Omega$ that were included in the stack. 
Uncertainties
in the stacked flux (tabulated in AB magnitudes) 
are mostly dominated by calibration systematics.}
\end{deluxetable}

\section{Stacking Analysis}

To establish the average IR intensity of faint optical sources that  
are unresolved in {\it Spitzer} IRAC images we stack
the IRAC images at the spatial coordinates of faint, optical sources which are 
detected in the GOODS optical data. The stacking begins by first masking all $>3\sigma$ detected
IRAC sources with $m_{\rm AB}$(3.6$\mu$m)$<26.7$ mag in the GOODS/IRAC mosaics\footnote{A description
of the GOODS mosaics can be found at http://ssc.spitzer.caltech.edu/legacy/}. This limit
is fainter than the 50\% completeness limit of 24.7 mag.
The mask has a radius of 13.5$\arcsec$ for a source with 18 mag
and scales linearly with magnitude down to 2.4$\arcsec$ for 
sources fainter than 22 mag.
We then identified sources in the GOODS ACS catalogs (Giavalisco et  
al. 2004) which are in unmasked
regions of the IRAC mosaics and are therefore IRAC undetected.
While there are close to $22,000$ ACS sources which are unmatched  
with IRAC sources to within $\sim0.5\arcsec$,
only 6,160 (GOODS-N) and 5,441 (GOODS-S) sources are sufficiently
separated from brighter IRAC sources so as to be in unmasked regions  
of the IRAC image. 
The IRAC images for each GOODS field were stacked at the pixel locations corresponding to  
these $\sim$6000 ACS sources. The stacking was undertaken by
making 12$\arcsec$ image cutouts from the masked IRAC mosaic, centered
on each ACS source position. The cutouts were then coadded using
a 3$\sigma$ clipped median resulting in a median stacked image
of the sources, as shown in Figure 1. 
A count of the number of sources which
contribute to the stacked flux is also maintained resulting in 
an effective exposure map.
Pixels which have been masked do not contribute either flux or exposure in the stacking
process.
Circular aperture photometry
was performed on this stacked image and the flux measured in a 3.6$\arcsec$
radius beam. The sky background was estimated from the stacked image
by measuring the sigma-clipped median of pixels within
a surrounding sky annulus spanning $5-6\arcsec$. 
The profile of the source had a full-width at half-maximum (FWHM) of 1.6$\arcsec$
and was consistent with an unresolved point source at the resolution
of {\it Spitzer}.
Aperture corrections were estimated by performing identical photometry on a calibration 
point source source and correspond to a factor of 1.14.
After background subtraction and aperture corrections to the photometry were performed,
the resultant stacked 3.6$\mu$m flux density is 130$\pm$1.3 nJy.

In order to assess the reliability of the stacked flux, we stacked
the IRAC images at an identical number of positions which are offset
from the nominal ACS source positions by random amounts, up to 9$\arcsec$
from the original source coordinates. This has the
advantage of accounting for any possible contamination of the stacked flux due to
the wings of nearby sources which may not be completely masked, while measuring
the same local sky background. The random stacks were repeated 100 times and the
flux of the stack measured. The standard deviation of these values
results in a flux density of 5 nJy which is a factor of 4 larger than the 
uncertainty simply due to the background noise term due to source confusion
and the low-level flux from the wings of masked sources.
Thus, the stacking results in a strong detection of these faint optical galaxies
and we adopt a stacked 3.6$\mu$m flux density of 130$\pm$5 nJy.

This procedure was repeated at each IRAC wavelength independently for each
GOODS field to assess the reliability of the result. Table~1 summarizes stacking the results for
all IRAC wavelengths between 3.6 $\mu$m and 8.0 $\mu$m.
Through the stacking analysis, we obtain a strong detection of these faint optical galaxies and
measure a flat flux spectrum (F$_{\nu}\propto\nu^{0.08\pm0.3}$). This is consistent with
the flat frequency spectrum of IR fluctuations between 3.6 $\mu$m and 8.0 $\mu$m
found by Kashlinsky et   al. (2005).

In order to estimate the total contribution to the IRB intensity of these sources,
we multiply the median stacked flux by the exposure map.
Based on the surface density of about 6000 galaxies in each GOODS field,
we establish the absolute 3.6 $\mu$m IR intensity of these
faint optical sources  to be 0.12$\pm$0.01 nW m$^{-2}$ sr$^{-1}$. We note that this 
estimate is larger than the value obtained by simply multiplying the quoted number of sources
($\sim$6000) with the median flux. This is because the number of sources measurement requires that
a source be unmasked out to a radius of 3.6$\arcsec$ while the exposure map takes into
account the exact number of sources which contribute to the stacked flux in each pixel.
Since some of the source may be partially masked within the 3.6$\arcsec$ radius,
the number of sources is smaller than the number of sources as estimated from the exposure map.

Furthermore, this estimate of 0.12 nW m$^{-2}$ sr$^{-1}$ only  
accounts for sources that remain outside the IRAC mask and are not confused with bright  
IRAC sources. Assuming the average flux from the stack also applies to
 sources affected by the mask, we estimate the absolute IRB intensity to be
as high as 0.35 nW m$^{-2}$ sr$^{-1}$. This is well within the range of  
0.1 to 0.8 nW m$^{-2}$ sr$^{-1}$ estimated for the contribution of faint sources
below the individual point source detection limit in IRAC images
through clustering models (Sullivan et al. 2007).

To further study the nature of these sources, beyond {\it Spitzer}/IRAC wavelengths, we adopted
a similar procedure and stacked the same population 
at other wavelengths at which the GOODS fields have imaging
data. Specifically, for GOODS-N, we utilize the KPNO U-band 
\citep{Capak04}, {\it Hubble}/ACS $BViz$ (Giavalisco et al. 2004), and
near-infrared HK' \citep{Capak04}. For GOODS-S, we use the CTIO U-band
and the ESO/ISAAC near-infrared data in the $JHK$ bands in addition to the
{\it Hubble} and {\it Spitzer} data. For the {\it Hubble} data, we simply
measure the median flux values of the sources since they are individually detected
at the ACS wavelengths. For the other wavelengths, we excised subimages for each source
and estimated a sigma clipped average of the subimages
to obtain the stacked flux, as was done for the IRAC data. Aperture corrections, background estimates
and noise calculations were performed as described for the IRAC data.

\begin{figure}[!t]
\centerline{\psfig{file=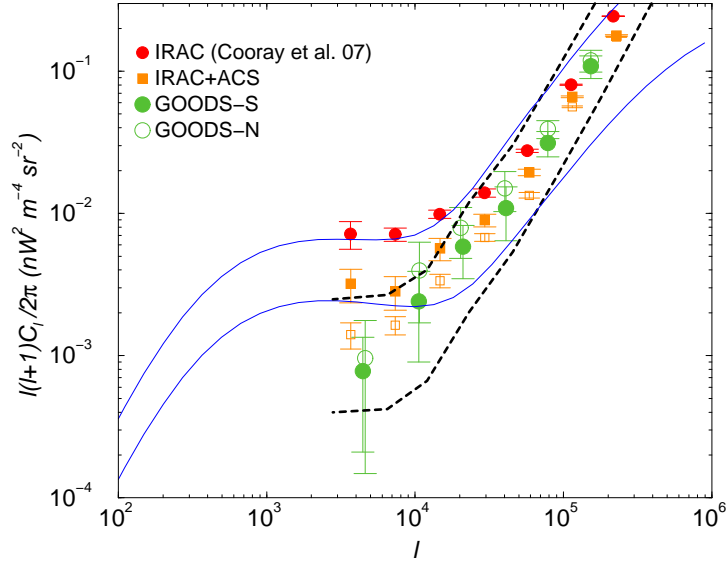,width=3.0in,angle=-90}}
\caption{
The top solid circles show the angular power   spectrum of unresolved
IR fluctuations in IRAC channel 1 (3.6 $\mu$m) of GOODS-N field. 
The solid squares show the angular power spectrum 
that masks the locations of faint ACS
sources that are undetected in IRAC.
The same power spectrum published in Cooray et al. 2007 is shown with open squares and the difference accounts for a new correction we have introduced here
for the window function of the ACS source mask.
With ACS sources masked, the fluctuation power spectrum amplitude is reduced 
at $\ell \sim 7 \times 10^3$ by a factor of about 2 ($\pm 0.6$). 
The circles and dashed lines below show the predicted power spectrum
for faint optical sources with their 3.6 $\mu$m number counts 
distributed with  two faint-end slopes:
the open and filled circles are for clustering in GOODS-N and GOODS-S assuming the steeper slope for counts
with $\alpha \sim 0.6$, while the dashed lines bracket the clustering  
for a flatter slope with $\alpha \sim 0$ ($dN/dm \propto m^{\alpha}$ when
$m > 26$)
For reference, the two solid lines are predictions from clustering models 
of the faint source population with the
 range covering the uncertainty in model parameters (Sullivan et al. 2007).}
\label{fig:irb}
\end{figure}

\section{Redshift and Stellar Mass Density Estimates}

In order to understand the physical properties of
these faint galaxies, we fit the stacked flux in Table 1 with Bruzual \& Charlot (2003; BC03) population synthesis models. Redshift,
mass, extinction, age and e-folding time of star-formation were left
as free parameters. The redshift range was $0.1-6$ in steps of 0.1, extinction range was A$_{\rm V}=0-5$ in steps of 0.1,
age $t$, varied from 10$^5$ yr to the age of the Universe at each redshift, while the e-folding time $\tau$ 
spanned the entire range from instantaneous to constant star-formation.
For GOODS-N, we find a best fit redshift of 1.9,
stellar mass of 1.3$\times10^{8}$~M$_{\sun}$, A$_{\rm V}$=1.4 mag, $t$=10 Myr and $\tau$=200 Myr, 
indicating an ongoing reddened starburst (Figure 2).
For GOODS-S, which has far better near-infrared data,
we find a best fit redshift of 2.1, stellar mass 
of 1.4$\times10^{8}$~M$_{\sun}$, A$_{\rm V}$=1.1 mag, $t$=15 Myr, $\tau$=200 Myr. 
The 90\% confidence interval for the redshifts as determined from the lowest chi-square fit
at each redshift, spans the range $1.9<z<2.9$ for GOODS-S and $1.3<z<2.7$ for GOODS-N.
The $U-$band detection implies that the majority of the sources must be at $z<2.5$.
The sharp increase in the flux in the IRAC channels compared to the NIR bands implies
that the 1.6$\mu$m bump must be in the IRAC passbands. This implies $z>1.3$. Thus, even a qualitative
comparison of the stacked flux densities with a typical galaxy SED results in a redshift range similar
to what we estimate from the chi-square values. We adopt $1.9<z<2.7$ as the redshift range of these sources.
The uncertainty in the derived stellar mass is about a factor of two, ranging
from $1.3-3.0\times10^{8}$~M$_{\sun}$.

For our best fit template, the rest-frame 1500\AA\ UV luminosity
of the stacked galaxy is 1.8$\times$10$^{9}$~L$_{\sun}$ while
its rest-frame V-band luminosity is 1.2$\times$10$^{9}$~L$_{\sun}$.
Since L$_{*,{\rm UV,z=3}}$=5.8$\times$10$^{10}$~L$_{\sun}$ and 
L$_{*,{\rm V,z=3}}$=8$\times$10$^{10}$~L$_{\sun}$ \citep{Steidel99, Mar07}, it implies that these
faint objects are galaxies which are about $\sim30-60$ times fainter than
the characteristic luminosity of field galaxies at these redshifts.

Although the exact redshift distribution of these faint galaxies
is not known at this time, it is illustrative to estimate
the fraction of the stellar mass density hidden in these objects.
We assume that the $\sim$6000 galaxies that we have stacked on
are distributed between $1.9<z<2.7$ with a stellar mass of 1.4$\times$10$^{8}$~M$_{\sun}$.
This corresponds to a co-moving stellar mass density of 2$\times$10$^{6}$~M$_{\sun}$\,Mpc$^{-3}$ which is a strong lower limit. Assuming that our
stacked flux is typical for the $\sim$22000 ACS sources which are unmatched
to IRAC sources, would imply a total stellar mass density in faint galaxies
of 6.7$\times$10$^{6}$~M$_{\sun}$\,Mpc$^{-3}$ at $z\sim2.4$.  For comparison, the stellar mass density in Lyman-break
galaxies at these redshifts is 3.6$\times$10$^{7}$~M$_{\sun}$\,Mpc$^{-3}$ \citep{MED03}.
Thus, the faint, sub-L$_{*}$ galaxy population could account for upto 20\% of the  total stellar mass density at $z \sim 2.4$.

\section{Expected Clustering}

We also measure the clustering of these sources to establish the amplitude of unresolved fluctuations at
IRAC bands. Since we are interested in the anisotropy of IR light,
to establish the power spectrum of fluctuations,
we need the flux distribution within this population. The stacking analysis only allows us to establish either the total
or the average flux of this sample. This forces us to make an estimate
of the flux distribution and use that to predict clustering of unresolved fluctuations  
produced by these sources. The angular power spectrum
of IR fluxes associated with these sources  is determined
using the same technique as described in Sullivan et al. (2007).  
We account for the flux distribution by randomly assigning IR flux to the $\sim$  
6000 optical source locations   between 70  nJy and 700 nJy such that the total flux assigned 
is the same as the total flux measured for these sources from the stack.
We assign fluxes such that the number counts trace the  extrapolated slope from known  IRAC
counts down to $m_{\rm AB}(3.6\mu m)$ of 24.7 mag. Given the uncertainty in the
faint-end slope, we take the slope $\alpha$ ($dN/dm \propto m^\alpha$) to be
either 0.6 or 0.0. We tested how our
predictions change with variations to these parameters
and found consistent results within errors. Since the flux assignment is 
random, to get an average of the expected clustering, we randomize the
assignments and the fluxes and use a Monte-Carlo approach to obtain  
the mean and variance of clustering.

We summarize our results in Figure~3, where we also
compare with a direct measurement of the clustering of IR  
fluctuations and expectations based on the clustering  models of Sullivan et al. (2007; solid lines). In Figure~3, we also show  the IR fluctuation spectrum when the faint optical
sources that we stack on are masked from the IRAC image with open squares
(Cooray et al. 2007). The solid squares show the revised measurement after correcting for a bias
associated with the window function introduced by the large mask.
To understand this correction noted in Kashlinsky et al. (2007a), 
we note that the fluctuation spectrum measurable in Fourier (multipole) space in the presence of a mask is
$\hat{C}_\ell = \sum _{\ell'} W_{\ell \ell'} C_{\ell'}$, where
$W_{\ell \ell'}$ is the window function associated with the mask and $C_{\ell'}$ is the power spectrum of interest. We recover the latter 
with measurements of $\hat{C}_{\ell}$  by first generating  $W_{\ell \ell'} $ associated with our mask (Appendix A of Hivon et al. 2002) 
and then iteratively inverting for $C_{\ell}$ using an inversion similar to
Dodelson \& Gazta\~naga (2000). The inversion agrees to 10\% with a separate
estimate of the angular power spectrum using a likelihood method where
one maximizes the likelihood of an estimated $C_\ell^{\rm est}$ with  a model spectrum $C_{\ell}^{\rm inp}$ 
using $C_\ell^{\rm est}=\sum_{\ell'} W_{\ell'} C_{\ell'}^{\rm inp}$ given the measurements
$\hat{C}_{\ell}$ and the error $\sigma_{\ell}$. This procedure can be described
as minimizing the $\chi^2$ with $\chi^2=(\hat{C}_{\ell}-C_{\ell}^{\rm est})^2/\sigma_{\ell}^2$.

As shown in Figure~3, the difference between the fluctuation amplitude of unresolved IR fluctuations
and the fluctuation amplitude with ACS sources masked is about a factor of 2 ($\pm$ 0.6) at $\ell \sim 7 \times 10^3$.
Our measurements then suggest that, from this difference, up to 50\% of IR fluctuations can be accounted by faint ACS sources.
It could be that by accounting for further fainter optically sources, this fraction can be increased, but
it is quite
unlikely that 100\% of IR fluctuations reported in Kashlinsky et al. (2005, 2007b) are generated by $z>6.5$ galaxies hosting mostly Pop~III stars
that are optically invisible.

The difference between IR fluctuations with and without ACS sources masked must be reproduced by the 
clustering of IR light produced by the same ACS sources. With flux assignments based on the
stacking analysis, we find 
that the difference is reproduced (though these indirect estimates of
the expected clustering IR fluctuations produced by faint ACS sources are uncertain due to uncertainties in the flux distribution).
We conclude that irrespective of whether we measure the power spectrum in
the IRB fluctuations after masking optical sources or whether we measure the power spectrum of faint optical
galaxies with an average flux based on our stacking analysis, we find that the power in faint ($z_{AB}\sim27$ mag), $z\sim2.5$, 
galaxies accounts for at least $\sim$30$-$50\% of the power in the IR background fluctuations
on angular scales of $\sim$4$\arcmin$, where the measurement is not strongly sensitive to cosmic variance. 

On similar angular scales, Kashlinsky et al. (2007a) find a power of 0.015 nW~m$^{-2}$~sr$^{-1}$ for the optical galaxies
with $z_{AB}>$26.6 mag in HDFN-E2, without a quantification of the uncertainty.  
They also demonstrate that the power increases as brighter galaxies are included. 40\% of the galaxies
that go into our stacking analysis
are brighter than this threshold and therefore the power they measure is a lower limit to our
power spectrum shown in Figure 3.
This can be compared to their measured value of 0.035 nW~m$^{-2}$~sr$^{-1}$
for the amplitude of power in the IR background fluctuations in the HDFN on the same angular 
scale/multipole moment.
Making use of their estimate of the fluctuations associated with faint, optical galaxies at low redshifts, we find that 
Kashlinsky et al. (2007a, 2007b) also find a $\sim$ 40\% contribution from such sources to
fluctuations at $\sim$ 4 $\arcmin$ angular scales, though their contribution is ignored in 
their subsequent interpretation
that assumes all fluctuations are from Pop~III stars before reionization.
The analysis of the optical sources in GOODS-S is not presented in Kashlinsky et al. (2007a). The 
fundamental discrepancy between our two groups
arises on larger angular scales, at 400 arcsec or larger, where they claim the power spectrum is rising. We are unable
to constrain the power spectrum on such large angular scales ($\ell \sim 1500$) due to the cosmic variance associated with the
limited size of the GOODS fields used in our analysis, which becomes important at angular scales above 400 arcseconds.

\section{Discussion}

The Kashlinsky et al. (2007c) interpretation for IR fluctuations involves
$z>6.5$ galaxies with an IR background intensity of $\gtrsim$ 1 nW m$^{-2}$  
sr$^{-1}$ at 3.6 $\mu$m. This is a large intensity given that all resolved sources, so far,  
lead to an IR background intensity between $\sim$ 6 nW m$^{-2}$ sr$^{-1}$ 
(Sullivan et al. 2007; Fazio et al. 2004) and $\sim$ 10 nW m$^{-2}$ sr$^{-1}$ (Levenson \& Wright 2008).  
To avoid individual  detections of a large number of $z>6.5$ galaxies, such a
high background intensity must then be hidden in a source population with  
a large surface density but with individual faint fluxes of around 10 nJy for  
each source. Given the large surface density of the   population, pixel to pixel intensity
variations  are a small fraction of the total intensity produced by  
those sources.

The alternative explanation is that a reasonable fraction of fluctuations arises from
a population of sources that have a low surface density but are just below
the {\it Spitzer}/IRAC detection threshold. The results from the stacking analysis presented here 
suggest that faint optical galaxies, with an average source flux of  
around 130 nJy at 3.6 $\mu$m, contribute between $\sim$30$-$50\% of the power
in the fluctuations on the largest angular scales that we are able
to measure.   The average IR absolute intensity produced by these sources is in the range
0.12-0.35 nW m$^{-2}$ sr$^{-1}$ at 3.6 $\mu$m, in agreement with the
Kashlinsky et al. (2005) estimate that the contribution
from faint extragalactic sources is $\sim$0.15 nW m$^{-2}$ sr$^{-1}$. 
However, they claim that such faint objects
cannot account for the strong clustering signal in unresolved IR light. We have demonstrated in this paper that a significant
fraction of the fluctuations do arise from these faint galaxies.
These sub-L$_{*}$ galaxies contribute up to 20\% of the stellar mass density at $z \sim 2.5$
and can be described as a low luminosity version of well-studied Lyman-break galaxies.

While these sources do not fully explain all IR fluctuations measured
with deep IRAC images, in Cooray et al.  
(2007), we placed a conservative upper limit that any
contribution from $z>6.5$ sources must have an absolute intensity less than  
about 0.6 nW m$^{-2}$ sr$^{-1}$.  Beyond {\it Spitzer} IRAC, studies have also been conducted at  
lower near-IR wavelengths with NICMOS. Though limited 
by the small field-of-view of NICMOS, they show that   IR fluctuations at 1.6 and 1.25 $\mu$m are
more consistent with a $z<8$ origin  than a high redshift  
interpretation (Thompson et al. 2007a, 2007b).
Beyond IR fluctuations, it will be interesting
to further study the nature of the faint, dwarf galaxy population at $z=2$ to 3 
to understand the contribution it makes towards bridging the systematic offset between the 
co-moving stellar mass density and the integrated star-formation rate density (Hopkins \& Beacom 2006).

{\it Acknowledgments:}
We thank Mark Dickinson for helpful suggestions. We also acknowledge the 
contributions of various members of the GOODS team who are associated with key
aspects of data processing and handling.
This research was funded by NASA grant NNX07AG43G, NSF CAREER grant AST-0645427,
award number 1310310 from {\it Spitzer} for Archival Research, 
and program number HST-AR-11241.01-A by NASA through
a grant from the Space Telescope Science Institute, which is operated  
by the Association of Universities for Research in Astronomy, Incorporated,  
under NASA contract NAS5-26555.

\end{document}